# Effect of Non-Stoichiometry on Magnetocaloric Properties of HoB$_2$ Gas-Atomized Particles

Takafumi D. Yamamoto[1], Hiroyuki Takeya[1], Pedro Baptista de Castro[1,2], Akiko T. Saito[1], Kensei Terashima[1], Takenori Numazawa[1], and Yoshihiko Takano[1,2]

[1]National Institute for Materials Science, Tsukuba, Ibaraki 305-0047, Japan
[2]University of Tsukuba, Tsukuba, Ibaraki 305-8577, Japan

We fabricate gas-atomized particles by inductively melting electrode rods of HoB$_{2-x}$ ($x = -0.3, 0, 0.3,$ and $1.0$) and investigate the effect of non-stoichiometry on the phase fraction, microstructure, and physical properties. Shifting the stoichiometric ratio of the electrode rod to the B (Ho)-rich side increases HoB$_4$ (Ho) phase in the resulting atomized particles. Even if the atomized particles contain 15–20 weight percent (wt.%) of the impurity phase, the influence of which on the physical properties is less severe: the maximum value of the magnetic entropy change is only reduced by 10% compared to HoB$_{2.0}$ particles. We further find that the ductile Ho phase exists so as to fill the space between the brittle HoB$_2$ phases in the atomized particles, which may be beneficial to the mechanical properties of the particles. Our findings suggest that it would be better to use the Ho-rich electrode rods than the stoichiometric ones to produce HoB$_{2-x}$ particles with more suitable properties as a magnetic refrigerant for magnetic refrigeration systems.

## I. Introduction

MAGNETIC REFRIGERATION is a cooling technology based on the magnetocaloric effect (MCE) of solid-state materials and has a broad range of potential applications such as domestic air conditioning and refrigerators [1-4], gas liquefaction [5,6], and space cryogenics [7]. Since this technology does not require hazardous/greenhouse refrigerant gases and gas-compressors, magnetic refrigeration systems can be environmentally friendly, compact, noiseless, and possibly more efficient than conventional gas-based refrigeration systems [8,9]. For developing high-performance refrigeration systems, one needs not only to explore for magnetocaloric materials with large MCEs [6, 10-12], but also to process the promising materials into magnetic refrigerants with appropriate shapes to each system. For instance, the materials should be shaped into spherical particles for use in the often-used systems using an active magnetic regenerator (AMR) cycle [13,14].

Recently, it was revealed that a rare-earth diboride HoB$_2$ is a giant magnetocaloric material for low temperature applications [15]. The magnetic entropy change $\Delta S_M$ takes the maximum value of 40.1 J kg$^{-1}$ K$^{-1}$ for a magnetic field change of 50 kOe near the Curie temperature $T_C$ of 15 K. Owing to the $T_C$ close to the liquid hydrogen temperature of 20.3 K, HoB$_2$ is expected as a promising material for use in magnetic refrigeration systems for hydrogen liquefaction. In the previous work [16], we succeeded in producing HoB$_2$ particles by an electrode induction melting gas atomization (EIGA) technique in which an electrode rod of the material is inductively melted in a non-contact manner. The resulting particles have good particle morphology, a sub-millimeter size suitable for AMR systems, and giant magnetocaloric effect as well as the ingot counterpart, suggesting their good potential as a magnetic refrigerant.

Nevertheless, it should be noticed in performing the EIGA process on HoB$_2$ that this material is a peritectic system. According to a B-Ho phase diagram [17] shown in Fig. 1, HoB$_2$ decomposes into a solid HoB$_4$ and a Ho-rich melt at 2200 °C, and then the system completely melts at 2350 °C. Due to such incongruent melting, the chemical composition of the molten metal may deviate from the nominal composition of HoB$_2$, leading to impurity phases such as HoB$_4$ and Ho in the resulting atomized particles. In this context, it would be meaningful to reveal how the impurity phases affect the magnetocaloric properties of HoB$_2$ atomized particles. For this purpose, we prepare gas-atomized spherical particles of HoB$_{2-x}$ ($x = -0.3, 0, 0.3,$ and $1.0$) by the EIGA technique and investigate the effect of non-stoichiometry on the phase fraction, microstructure, and physical properties.

## II. Experimental Methods

Electrode rods of HoB$_{2-x}$ ($x = -0.3, 0, 0.3,$ and $1.0$) were prepared by arc-melting in an Ar-atmosphere from Ho (3N) and B (3N) elements with composition ratios corresponding to each $x$-value. Each rod was melted repeatedly for homogenization. In the EIGA process, each electrode rod, fixed and dangled by a chuck at one end, was inductively melted at the other end by an induction coil in an Ar-atmosphere. After the molten metal freely fell into the lower sample chamber, it was atomized by jetting Ar-gas and solidified into particles while flying in the chamber. The collected atomized powder was sieved and divided into spherical particles and irregular ones, respectively. The details of preparation process of spherical particles are described elsewhere [16].

Powder X-ray diffraction (XRD) measurements were performed at room temperature by a Rigaku X-ray diffractometer with Cu $K\alpha$ radiation for phase identification. The microstructure and element distribution of the atomized particles were observed by using a Hitachi TM-4000 scanning electron microscope (SEM) coupled with X-ray energy-dispersive spectroscopy operated at 10 kV. For the cross-sectional observation, the spherical particles were embedded in Bakelite resin containing carbon filler and polished to a flat surface. Magnetization ($M$) measurements were carried out by a Quantum Design MPMS. Specific heat ($C_p$) in zero magnetic field was measured using a thermal relaxation method by a Quantum Design PPMS. These measurements were performed on a dozen or so spherical particles. The same sample was used in both magnetization and specific heat measurements. The magnetic entropy change ($\Delta S_M$) was calculated from a series of the iso-field magnetization ($M$-$T$) curves measured under various magnetic fields ($H$) between 0.1 kOe and 50 kOe by using one of the Maxwell's relations,

$$\Delta S_M(T, \Delta H) = \int_0^H \left(\partial M / \partial T\right)_H dH \quad (1)$$

where $\Delta H$ is the magnetic field change from zero to $H$.

## III. RESULTS AND DISCUSSION

### A. Phase fraction

Fig. 2 shows the powder XRD patterns of HoB$_{2-x}$ particles with 355–500 μm diameter. The Bragg peak positions calculated from the AlB$_2$-type hexagonal structure of HoB$_2$ are also shown at the bottom. The major peaks are indexed by HoB$_2$ in all the samples, though the relative peak intensity is weakened for $x = 1.0$. No shift of the HoB$_2$ peaks with $x$ is observed, indicating no effect of non-stoichiometry on the lattice constant of the HoB$_2$ phase. The HoB$_4$ phase stands out as a secondary phase on the B-rich side, while the presence of the Ho phase becomes prominent on the Ho-rich side. This trend is in line with what expected from the phase diagram.

For evaluating the phase fraction of each sample, the Rietveld analysis of the XRD patterns were carried out [18-20]. The results are summarized in Table I. The stoichiometric sample is mostly in the HoB$_2$ phase, but a bit trace (5.2 wt.%) of the HoB$_4$ phase is detected. When the absolute value of $x$ is 0.3, the atomized particles contain 80–85 wt.% HoB$_2$ phase and 15–20 wt.% impurity phase. The fraction of the HoB$_2$ phase is largely decreased for HoB$_{1.0}$ particles, with about 38 wt.% Ho phase present.

### B. Microstructures

Figs. 3(a)-3(d) show SEM backscattered electron images of magnified cross section of HoB$_{2-x}$ particles with 355–500 μm diameter. The figures are arranged from (a) to (d) in ascending order of the Ho content in Table I. In the HoB$_{2.0}$ particles (Fig. 3(b)), the HoB$_2$ phase (bright region) occupies most of the region and the minority HoB$_4$ phase (dark region) also exists, being consistent with the XRD result. On the B-rich side, the HoB$_4$ phase increases and forms line-shaped domains. On the Ho-rich side, the areas of the HoB$_2$ phase reduce so that they eventually form a dendritic structure, which is seen as the bright region in Fig. 3(d). In parallel with this, the Ho phase appears in the atomized particles, and the areas expand so as to fill the space between the HoB$_2$ phases.

We notice that the Ho phase looks darker than the HoB$_2$ phase, even though the former should be brighter in the backscattered electron image. This is possibly because the exposed Ho surface was oxidized between polishing and observation. The small white spots in Fig. 3(d) seem to correspond to unoxidized Ho phase.

Meanwhile, one can find that the voids become less noticeable on the Ho-rich side. This is also evident from the cross-sectional images of the whole of the particle shown in Figs. 3(e)-3(h). Previously, we attributed them to the crystallographic nature of HoB$_2$, in other words, the ease of crystal orientation of this material [16]. Since the disappearance of these voids, however, appears to be linked to the increase of the ductile Ho phase, they are likely due to extrinsic factors such as friction against the brittle HoB$_2$ phase during polishing.

### C. Thermodynamic properties

Figs. 4(a)-4(c) show various thermodynamic properties of HoB$_{2-x}$ particles, where each thermodynamic quantity is given in the magnitude per total mass of the particles. First, let us review the HoB$_{2.0}$ particles. As shown in Fig. 4(a), the inverse magnetic susceptibility $H/M$ at 0.1 kOe above 200 K follows the Curie-Weiss law, $H/M = (T-\Theta_p)/C$, with the Curie constant $C$ of 0.08170 emu K$^{-1}$ g$^{-1}$ and the asymptotic Curie temperature $\Theta_p$ of 19.08 K. From the value of $C$, we obtain the effective magnetic moment $\mu_{eff}$ to be 11.04 $\mu_B$ per formula unit, which is close to the theoretical value for free Ho$^{3+}$ ions, 10.6 $\mu_B$ per formula unit. The value of $\Theta_p$ is nearly consistent with that reported in the literature [21]. Note that $\mu_{eff}$ can be converted from $C$ only if the molar weight can be defined properly, in that, if the sample is (mostly) single phase such as the HoB$_{2.0}$ sample. At low temperatures, $M$-$T$ curve at 0.1 kOe exhibits a rapid increase below 20 K and a kink anomaly around 11 K (Fig. 4(b)), corresponding to the ferromagnetic transition at 15 K and an unexplained magnetic transition at 11 K, respectively. These magnetic phase transitions also manifest themselves in $C_p$ as clear peaks, as can be seen in Fig. 4(c).

For $x = -0.3$, $H/M$ above 200 K can be described by the Curie-Weiss law with the $C$ of 0.07692 emu K$^{-1}$ g$^{-1}$ and $\Theta_p$ of 13.53 K. The smaller $\Theta_p$ possibly reflects the presence of the HoB$_4$ phase. HoB$_4$ is known to have $\Theta_p$ of −14.0 K [21] and exhibits successive antiferromagnetic transitions at $T = 7.1$ K and $T = 5.7$ K [22, 23]. A trace of these transitions shows up in thermodynamic properties below 10 K: a small hump of the $M$-$T$ curve (the inset of Fig. 4(b)) and a shoulder structure of $C_p$. Apart from them, however, the thermodynamic properties below 50 K are similar between $x = -0.3$ and $x = 0$, suggesting that they are mainly determined by the HoB$_2$ phase.

For $x = 1.0$, the linear temperature dependence of $H/M$ is also observed above 200 K, but an extrapolation of the linear part to $H/M = 0$ gives the higher $\Theta_p$ of 67.02 K. Moreover, a magnetic anomaly is found at around 125 K. These results imply that the strong influence of Ho, which has an $\Theta_p$ of about 87 K and an antiferromagnetic transition at 133 K [24, 25]. Actually, there are distinct differences in thermodynamic properties at low temperatures for $x = 1.0$ compared to $x = 0$: (i) a broad hump of the $M$-$T$ curve between 30 K and 50 K, (ii) weakening of the two specific heat peaks, (iii) a significant increase in $C_p$ above 20 K. The features (i) and (iii) can be explained by the thermodynamic properties of the Ho phase, the details of which are described in the Appendix. Regarding the feature (ii), since the heights of the two specific heat peaks themselves become lower, the Ho phase seems to essentially affect the physical properties of the HoB$_2$ phase in the HoB$_{1.0}$ particles.

For $x = 0.3$, $H/M$ bends downwards above 200 K and no longer follows the Curie-Weiss law. This result may reflect that the magnetism of the Ho and $HoB_2$ phases are intricately intertwined at high temperatures, rather than simply superposed. On the other hand, the $M$-$T$ curve and specific heat below 20 K are similar to those of $HoB_{2.0}$ particles. Furthermore, a slight increase in $M$ and $C_p$ are found between 20 K and 50 K, which is a trace of the Ho phase as described above. These results indicate that the thermodynamic properties of $HoB_{1.7}$ particles at low temperatures can be divided into two contributions from the Ho phase and the $HoB_2$ phase.

### D. Magnetocaloric properties

Fig. 5 shows the temperature dependence of $\Delta S_M$ for $\Delta H$ = 50 kOe in $HoB_{2-x}$ particles with 355–500 μm diameter. $\Delta S_M$ at $x = 0$ exhibits two peaks around 11 K and 15 K, with the maximum value of 39.2 J K$^{-1}$ kg$^{-1}$. In the $HoB_{1.0}$ particles, $\Delta S_M$ changes dramatically in both magnitude and shape. The maximum value of $\Delta S_M$ is decreased by about half compared to the $HoB_{2.0}$ particles. Moreover, the peak structure around 11 K changes into a shoulder structure. In addition, one can find a hump at around 30 K and the following tail that extends up to 50 K. These features can be ascribed to the Ho phase with the weight fraction of 38%, as is deduced from $\Delta S_M$ of a polycrystalline Ho shown in Fig. 5, for which the data are scaled with respect to the weight fraction.

In contrast to $x = 1.0$, the influence of the impurity phases on $\Delta S_M$ is less severe for $x = -0.3$ and $x = 0.3$. Although the decrease around 11 K is a little remarkable, the maximum value of $\Delta S_M$ around 15 K is reduced by only about 10% compared to the $HoB_{2.0}$ particles. At higher temperatures, $\Delta S_M$ at $x = 0.3$ is slightly predominant over $\Delta S_M$ at $x = -0.3$, owing to the contribution of the Ho phase to $\Delta S_M$. In this respect, the $HoB_{1.7}$ particles are better than the $HoB_{2.3}$ particles as a magnetic refrigerant. No trace of the $HoB_4$ phase in $\Delta S_M$ data is probably because the contribution to total $\Delta S_M$ for $\Delta H$ = 50 kOe is small, which is estimated to be 1–2 J K$^{-1}$ kg$^{-1}$ in magnitude below 7 K when considering the weight fraction and the expected value of polycrystalline $HoB_4$ itself [26].

### E. Effects of non-stoichiometry on $HoB_2$ particles

Shifting the stoichiometric ratio of the $HoB_{2-x}$ electrode rod increases the impurity phase in the resulting gas-atomized particles: the $HoB_4$ phase on the B-rich side and the Ho phase on the Ho-rich side. Nevertheless, our results demonstrate that the reduction in $\Delta S_M$ is minimized when the amount of the impurity phase is not so large. From Fig. 5 and Table I, it is presumed that even if the particles contain a little more impurity phase, e.g., 25 wt.% for $HoB_4$ phase and 20 wt.% for Ho phase, $\Delta S_M$ keeps the maximum value of about 30 J K$^{-1}$ kg$^{-1}$. The value is still as large as those in other promising magnetocaloric materials with similar $T_C$, such as $GdCoC_2$ [27] and HoN [28]. Thus, it is likely that some changes in the chemical composition of the molten metal of $HoB_2$ during the atomization process do not pose a problem for the magnetocaloric properties of the gas-atomized particles.

Another implication from the results is that the Ho phase in the $HoB_{2-x}$ atomized particles may be beneficial from a practical point of view. The disadvantage of $HoB_2$ as a magnetic refrigerant is that the brittleness of this material may cause the refrigerant particles to be broken by friction and the like during long-term use in AMR systems. The voids seen in the SEM images are probably due to the brittle nature. The disappearance of these voids with increasing the Ho content suggests that the ductile Ho phase works to prevent the expansion of cracks by filling the space between the $HoB_2$ phases. Accordingly, it is expected that mechanical properties of $HoB_2$ particles can be improved by the presence of the Ho phase.

Taken together the above, it can be said that an appropriate amount of the Ho phase may help to achieve both large magnetocaloric properties and good mechanical properties of the atomized particles. Therefore, we conclude that it would be better to use Ho-rich electrode rods rather than stoichiometric ones to produce $HoB_{2-x}$ particles more suitable as a magnetic refrigerant for magnetic refrigeration systems.

### APPENDIX

Holmium is known to has a complex magnetic phase diagram. In zero magnetic field, a phase transition from a paramagnetic to a basal-plane helical antiferromagnetic state is observed at ~133 K [29]. At 20 K, a first-order phase transition to a ferromagnetic state occurs, and a conical spiral spin structure with a small ferromagnetic moment along the $c$-axis is formed [29]. Furthermore, Ali et al. [30] have observed anomalies in the $M$-$T$ curves at 21, 42, and 98 K in low magnetic fields, corresponding to the transitions to the so-called spin-slip phases with different magnetic spiral wavevectors [31-33]. In the higher magnetic fields, other magnetic phases, such as helifan and fan phases, have been found in the wide temperature range between 20 K and 133 K [34, 35].

In this study, we measured the magnetization and the specific heat of a polycrystalline Ho as a reference sample. Fig. A1(a) shows the $M$-$T$ curves at 0.1 kOe in two different processes. In the process 1, a series of $M$-$T$ curves was first measured below 50 K under some magnetic fields from 50 kOe to 0.5 kOe, then the magnetization was measured after setting the magnetic field to 0.1 kOe at 50 K. In the process 2, the sample was cooled from 150 K to 50 K at 0 kOe and the magnetization was measured at 0.1 kOe below 50 K. The $M$-$T$ curve in the process 1, corresponding to those in Fig. 4(b), exhibits a broad hump between 30 K and 50 K, in addition to a rapid increase below 20 K derived from the ferromagnetic transition. As far as we know, there are no reports of the hump structure in the literatures. It might be associated with the history of magnetic field and temperature experienced by the sample before the measurement at 0.1 kOe. This speculation is also supported by the fact that the $M$-$T$ curve in the process 2 has no hump structure. Further investigation is needed for a complete understanding, but this hump structure is does found in Fig.4(b). Note that the $\Delta S_M$ shown in Fig. 5 was calculated from a series of $M$-$T$ curves in the process 1.

Fig. A1(b) shows the temperature dependence of the specific heat at 0 kOe in the polycrystalline Ho and the $HoB_{2.0}$ particles. The former $C_p$ is superior to the latter $C_p$ above 20 K, which is the cause of the significant increase in $C_p$ observed in the $HoB_{1.0}$ particles. In the polycrystalline Ho, no specific heat peak is found at around the ferromagnetic transition temperature of 20 K. This may be because the specific heat peak accompanying with a first-order transition is difficult to measure in the thermal relaxation method. It may also be related that a polycrystalline sample is used.

ACKNOWLEDGMENT

This work was supported by JST-Mirai Program, Japan (Grant Number JPMJMI18A3).

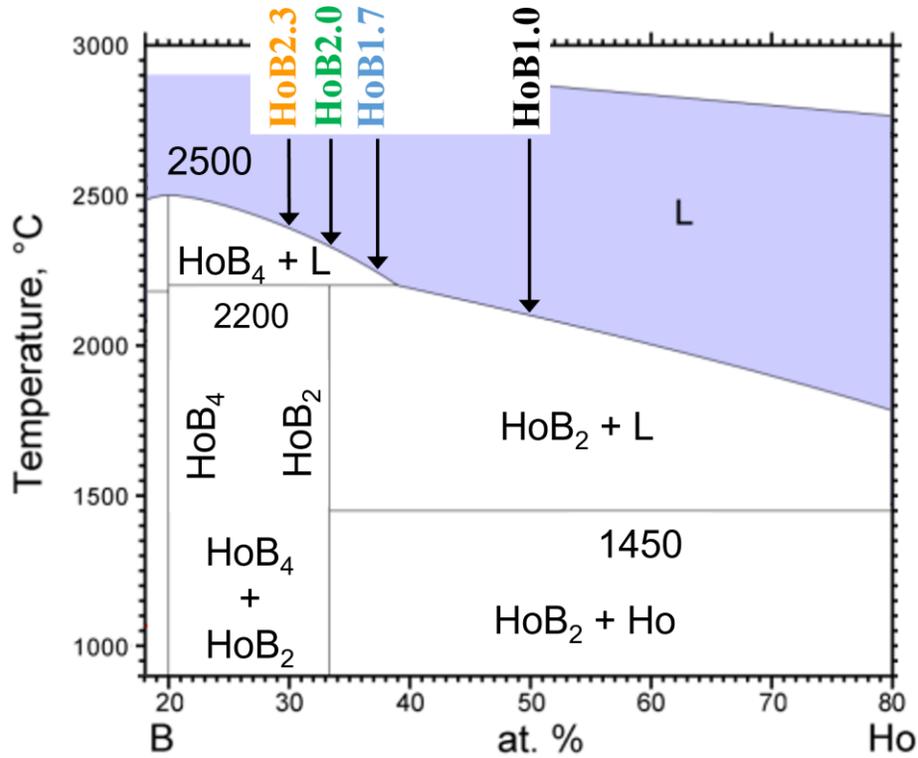

Fig. 1. (Color Online) Part of B-Ho phase diagram made from Ref. [17] in which the nominal composition of the samples prepared in this study is indicated.

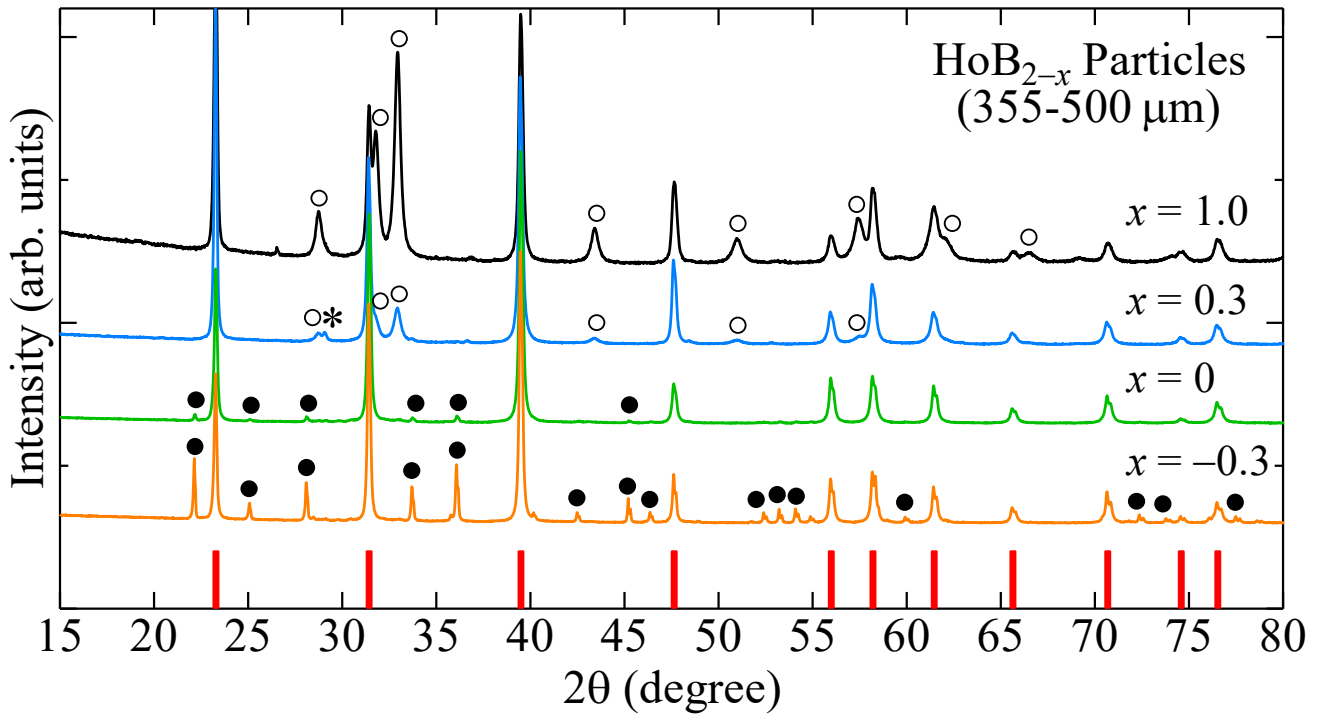

Fig. 2. (Color Online) Powder XRD patterns of $HoB_{2-x}$ particles with 355–500 μm diameter. The marks indicated the Bragg peaks of Ho (○), $HoB_4$ (●), and $Ho_2O_3$ (*). The bars at the bottom represent the Bragg peak positions of $HoB_2$.

TABLE I
PHASE FRACTION OF $HoB_{2-x}$ PARTICLES

| Nominal composition | $HoB_2$ (wt.%) | Ho (wt.%) | $HoB_4$ (wt.%) | $Ho_2O_3$ (wt.%) |
|---|---|---|---|---|
| $HoB_{1.0}$ ($x = 1.0$) | 62.1 | 37.9 | 0 | 0 |
| $HoB_{1.7}$ ($x = 0.3$) | 83.7 | 15.3 | 0 | 1.0 |
| $HoB_{2.0}$ ($x = 0$) | 94.8 | 0 | 5.2 | 0 |
| $HoB_{2.3}$ ($x = -0.3$) | 78.3 | 0 | 21.7 | 0 |

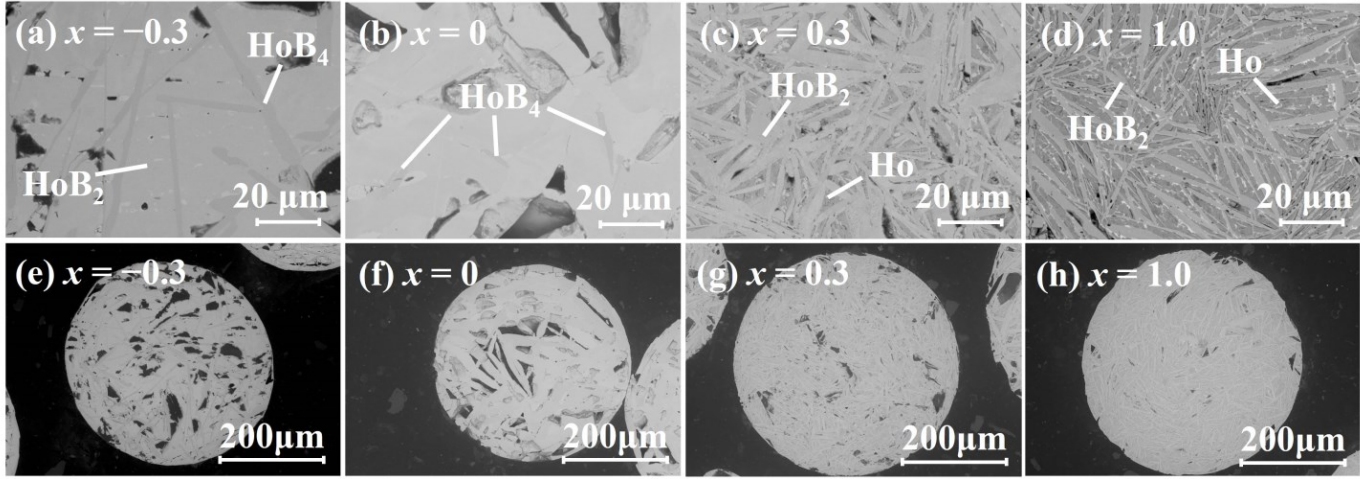

Fig. 3. SEM backscattered electron images of magnified cross section (a)-(d) and the corresponding the whole of $HoB_{2-x}$ particle (e)-(h) with 355–500 μm diameter.

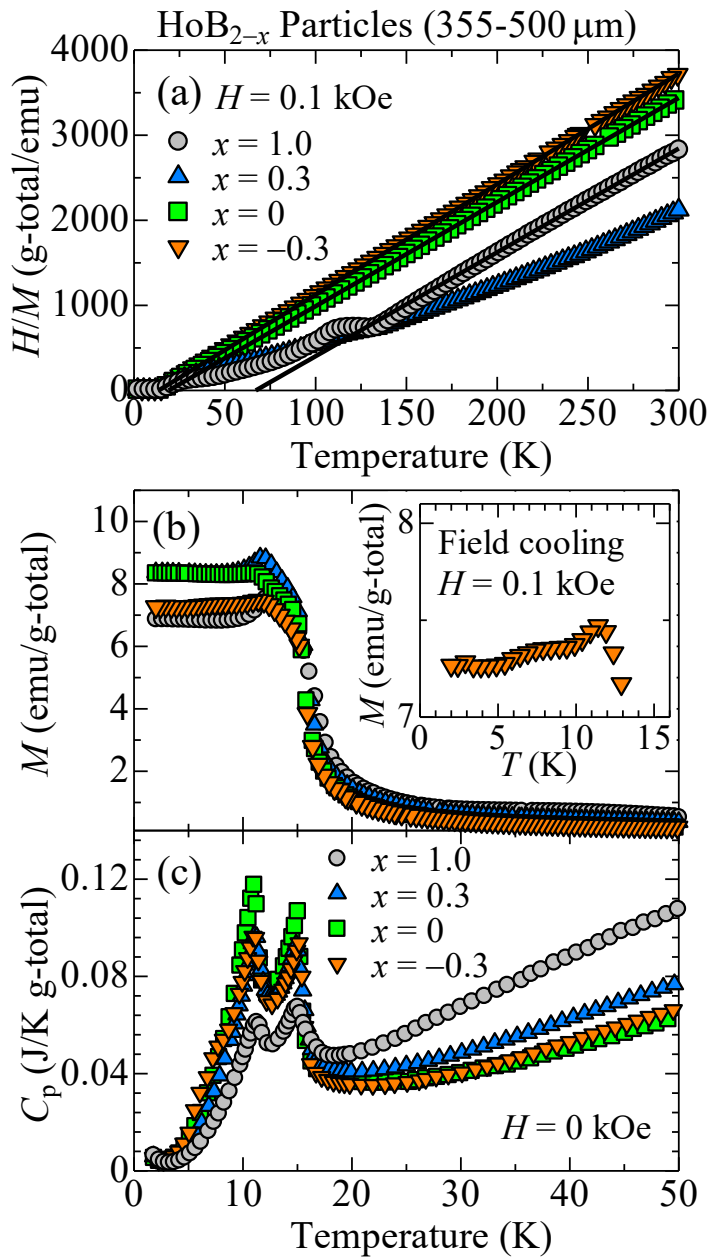

Fig. 4. (Color Online) Thermodynamic properties of HoB$_{2-x}$ particles. (a) Temperature dependence of inverse magnetic susceptibility $H/M$ at 0.1 kOe. The solid lines represent an extrapolation of the linear part to $H/M = 0$ (see text). (b) $M$-$T$ curves below 50 K measured in field cooling processes at 0.1 kOe. The inset shows an expanded view for $x = -0.3$. (c) Specific heat $C_p$ at 0 kOe as a function of temperature.

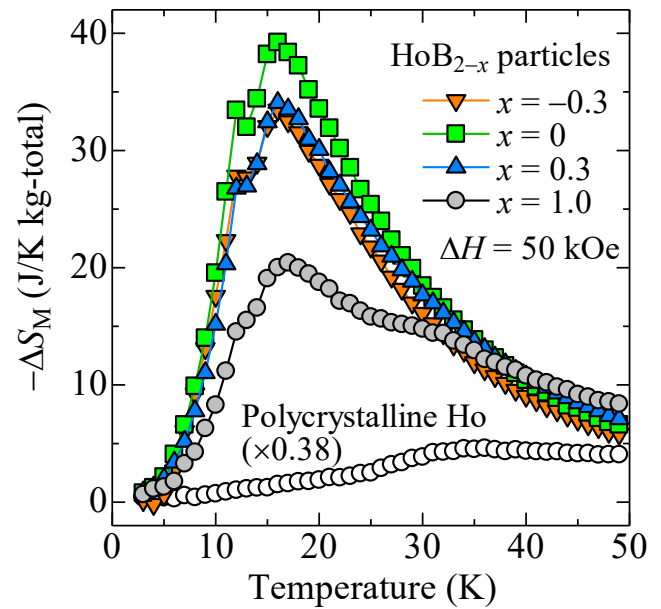

Fig. 5. (Color Online) Magnetic entropy change $\Delta S_M$ for the magnetic field change $\Delta H = 50$ kOe in the HoB$_{2-x}$ particles. The data for a polycrystalline Ho is also shown for comparison.

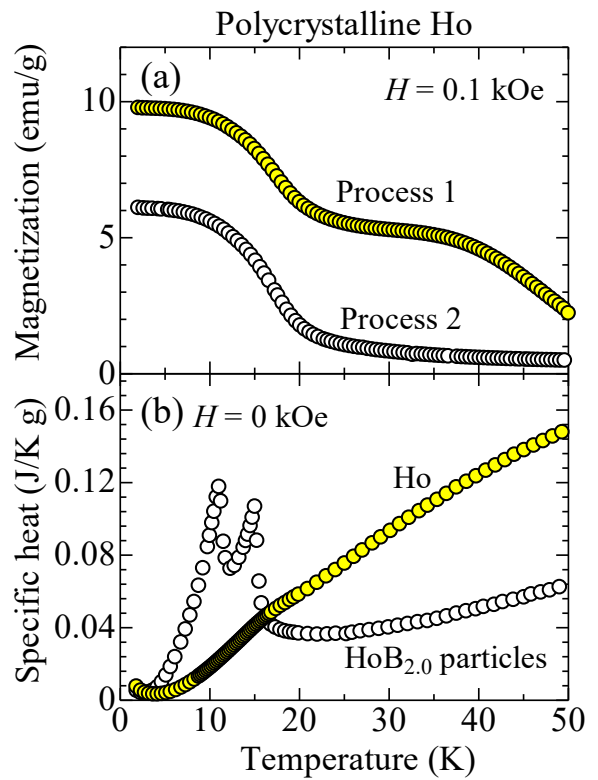

Fig. A1. (Color Online) Thermodynamic properties of the polycrystalline Ho. (a) Temperature dependence of magnetization at 0.1 kOe in different processes (see the Appendix). (b) Specific heat at 0 kOe as a function of temperature. The data for the HoB$_{2.0}$ particles are also shown for comparison.